\documentstyle[12pt,epsf,epsfig]{article}
\date{April 4, 1997}
\newcommand{\be}{\begin{equation}}
\newcommand{\ee}{\end{equation}}
\newcommand{\bea}{\begin{eqnarray}}
\newcommand{\eea}{\end{eqnarray}}
\newcommand{\mb}[1]{\mbox{\normalsize\boldmath $#1$}}
\newcommand{\uno}{{ 1\:\!\!\!\mbox{I}}}
\def\diag{\mathop{\mbox{diag}}}

\def\Tr{\mathop{\mbox{Tr}}}

\newcommand{\nn}{\nonumber}

\begin{document}
\title{ 
Flavour changing top decays in supersymmetric extensions 
of the standard model
}

\author{G.\ M.\ de Divitiis,  R.\ Petronzio and L.\ Silvestrini\\
\small Dipartimento di Fisica, Universit\`a di Roma {\em Tor Vergata} \\
\small and \\
\small INFN, Sezione di Roma II \\
\small Via della Ricerca Scientifica 1, 00133 Roma, Italy \\
\medskip
}

\maketitle

\begin{abstract} 
Flavour changing top decays $t \rightarrow c Z^0$, $t \rightarrow c g$ and
$t \rightarrow c \gamma$
are predicted with invisible rates within the standard model and may 
represent a window on new physics. We consider these processes in 
supersymmetric extensions of the standard model and we show that 
observable rates can be obtained only if the SUSY breaking is non universal and
flavour dependent.
 \end{abstract}

\vfill

\begin{flushright}
  {\bf ROM2F-97-08 ~~~~}\\
\end{flushright}

\newpage

\section{Introduction} 
Flavour changing neutral currents (FCNC) in the Standard Model (SM) are 
absent 
at tree level and suppressed by the GIM mechanism at 
one loop. 
They are particularly sensitive to large mass splittings between 
quarks of different generations.\\
Within the SM, the absence of tree-level 
FCNC implies that inside the loop only charged currents can
mediate the flavour change and therefore 
the large splitting of the third generation 
in the up sector can only be effective in processes with 
external {\em d\/}-type quarks.\\
The smallness of the mixing angles between the first 
two generations and the third one implies that the effects of a heavy top 
are 
particularly evident only in FCNC processes involving $b$ quarks, such as 
$B-\bar{B}$ mixing, radiative $B$ decays, etc.  

The extensions of the standard model to multi-Higgs doublet 
schemes allow for the presence of tree level FCNC which can be tested 
also with external up-type quarks and in particular in the 
FCNC top decays.
This study was pursued in 
refs.~\cite{2HDM},
 where it was shown that 
one could achieve values of the branching ratios of the
order of $10^{-6}$ , i.e. orders of magnitude bigger than the
standard model estimates of the
order of $10^{-11}$.

Tree-level 
FCNC vertices are also present in supersymmetric extensions of the SM. In 
particular, in the Minimal Supersymmetric Standard Model (MSSM), i.e. in 
the supersymmetric extension of the SM with minimal particle content, FCNC 
vertices involving fermions, sfermions and gauginos arise because of the 
misalignment of fermion and sfermion low-energy mass matrices. By imposing R 
parity, these vertices can only contribute to FCNC quark decays at one 
loop 
level via penguin diagrams in which the intermediate 
particles are gluinos\footnote{We will neglect neutralino 
contributions, because they are in general suppressed by powers of 
$\alpha_{W}/\alpha_{s}$.}. The difference with SM 
W-mediated penguins, and with SUSY chargino-mediated penguins, is that 
gluino-mediated penguins involving external up-type quarks are 
proportional to the mass splitting between {\em up-type\/} squarks which 
closely follows the one of the corresponding
non supersymmetric partners.

The purpose of this paper is to analyse neutral current
top decays, $t \rightarrow c Z^0$, $t \rightarrow c g$ and
$t \rightarrow c \gamma$, 
in supersymmetric extensions of the standard model.\\
We give the estimates for the
MSSM with flavour-universal soft SUSY breaking terms at the GUT scale, and 
also for a more general class of models where such a flavour universality 
does 
not hold. 
The universal case was already discussed in the literature 
\cite{Li_et_al,Couture_et_al}. In this case we have recomputed
both charginos and gluinos contributions.\\
We disagree with the results of refs.~\cite{Li_et_al,Couture_et_al} 
mainly in the evaluation of the relevant SUSY 
mixing angles and in a detail of the actual calculation of the decay amplitude.\\
We find that the SUSY mixing angle between the second and the third 
generation has been over-estimated by at least one order of magnitude.
We also find a difference in the result for the amplitude 
which can be traced back to the omission 
in previous papers of the diagrams involving a helicity flip in the 
gluino line, which dominate the branching ratios
when the gluino mass gets large.\\
The signal for the case where the SUSY breaking is universal
 turns out too low to be detectable. \\
In the non-universal case, where the soft SUSY breaking is not
flavour blind, the branching ratio can be as high as $10^{-5}$, which is 
maybe detectable. In this case the 
contributions from intermediate chargino exchange with respect to gluino
exchange can be generally neglected.
In Section 2 we review briefly those aspects of the MSSM which are relevant
for FCNC. In
section 3 we present the details of the calculation. The last section is
devoted to the presentation of the results and to the conclusions.

\section{The MSSM }

The superpotential of the MSSM is given by:
\begin{equation}
    W = Y^{u}_{ij}Q_{i}H_{2}U^{c}_{j}+Y^{d}_{ij}Q_{i}H_{1}D^{c}_{j}+
    Y^{\ell}_{ij}L_{i}H_{1}E^{c}_{j}- \mu H_{1}H_{2},
	\label{superp}
\end{equation}
which is the R-parity conserving supersymmetric generalization of the 
Yukawa interactions of quarks and Higgs bosons in the SM.

The breaking of SUSY is accounted for by introducing at the GUT scale 
$M_G$ the following soft SUSY breaking terms:
\begin{eqnarray}
	-{\cal L}_{s}\left(M_{G}\right) & = & 
		\widetilde{Q}_{L}^{i^{\dagger}}
	\left(\bar{m}^{2}_{\widetilde{Q}_{L}}\right)_{ij} 
	\widetilde{Q}^{j}_{L} +
	\widetilde{u}_{R}^{i^{\dagger}}
	\left(\bar{m}^{2}_{\widetilde{u}_{R}}\right)_{ij} 
	\widetilde{u}^{j}_{R} +
	\widetilde{d}_{R}^{i^{\dagger}}
	\left(\bar{m}^{2}_{\widetilde{d}_{R}}\right)_{ij} 
	\widetilde{d}^{j}_{R} 
	\nn \\
		& + & \widetilde{L}_{L}^{i^{\dagger}}
	\left(\bar{m}^{2}_{\widetilde{L}_{L}}\right)_{ij} 
	\widetilde{L}^{j}_{L} +
	\widetilde{\ell}_{R}^{i^{\dagger}}
	\left(\bar{m}^{2}_{\widetilde{\ell}_{R}}\right)_{ij} 
	\widetilde{\ell}^{j}_{R} + 
	\bar{m}^{2}_{1}h_{1}^{\dagger}h_{1} + 
	\bar{m}^{2}_{2}
	 h_{2}^{\dagger}h_{2} 
	 \nn \\
	 & + & 
	 \biggl[\bar{A}^{u}_{ij}\widetilde{Q}^{i}_{L}\tilde{u}^{j}_{R}
	 h_{2} +
	 \bar{A}^{d}_{ij}\widetilde{Q}^{i}_{L}\tilde{d}^{j}_{R}
	 h_{1} +
	 \bar{A}^{\ell}_{ij}\widetilde{L}^{i}_{L}\tilde{\ell}^{j}_{R}
	 h_{1} +
	 B\mu h_{1}h_{2} 
	\nonumber \\
	 &\,& +	\,\frac{1}{2}\left(
	 \bar{m}_{\tilde{g}}\tilde{g}^{T}C\tilde{g}+
	 \bar{m}_{\widetilde{W}}\widetilde{W}^{T}C\widetilde{W} +
	 \bar{m}_{\widetilde{B}}\widetilde{B}^{T}C\widetilde{B} \right) + 
	 {\rm h. \; c.}\biggr],
	\label{softgen}
\end{eqnarray}
where C is the charge conjugation matrix and $i,j=1,2,3$.  

We first consider a constrained version of the MSSM in which we impose 
universality of the soft breaking terms at the GUT scale:
\begin{eqnarray}
	&\,&	\left(\bar{m}^{2}_{\widetilde{Q}_{L}}\right)_{ij}=
		\left(\bar{m}^{2}_{\widetilde{u}_{R}}\right)_{ij}=
		\left(\bar{m}^{2}_{\widetilde{d}_{R}}\right)_{ij}=
		\left(\bar{m}^{2}_{\widetilde{L}_{L}}\right)_{ij}=
		\left(\bar{m}^{2}_{\widetilde{\ell}_{R}}\right)_{ij}=
		m^{2}_{0}\delta_{ij},
		\nn
	\\
	&\,&	\bar{A}^{u}_{ij}=A_{0}Y^{u}_{ij}\qquad
		\bar{A}^{d}_{ij}=A_{0}Y^{d}_{ij}\qquad
		\bar{A}^{\ell}_{ij}=A_{0}Y^{\ell}_{ij},
		\nn
	\\
	&\,&
		\bar{m}_{\tilde{g}}=\bar{m}_{\widetilde{W}}=
		\bar{m}_{\widetilde{B}}=m_{1/2}.
	\label{univ}
\end{eqnarray}
With these soft breaking terms, the model is defined by six parameters:
\begin{equation}
	m_{0},\quad m_{1/2},\quad A_{0},\quad \tan \beta,\quad \mu \quad
	{\rm and} \quad B.
	\label{sixpar}
\end{equation}
The request of radiative breaking of the electroweak symmetry 
reduces the parameters of the model to five:
\begin{equation}
	m_{0},\quad m_{1/2},\quad A_{0},\quad \tan \beta \quad {\rm and}\quad 
	{\rm sign}\,(\mu). 
	\label{fivepar}
\end{equation}

The low-energy values of the couplings in eqs.~(\ref{superp}) and 
(\ref{softgen}) are obtained by solving the following renormalization 
group 
equations \cite{Derendinger_et_al}-\cite{Bertolini_et_al}:
\begin{eqnarray}
	\frac{{\rm d} \mb{Y}^{u}}{\rm{d}t}& = & 
	-\frac{1}{2}\left[3\left(\mb{Y}^{u}\mb{Y}^{u^{\dagger}}+\Tr 
	\mb{Y}^{u}\mb{Y}^{u^{\dagger}}\right) + \mb{Y}^{d}\mb{Y}^{d^{\dagger}} - 
2C^{{\rm 
	u}}_{i}g^{2}_{i}\right] \mb{Y}^{u},
	\nonumber \\
	\frac{{\rm d} \mb{A}^{u}}{\rm{d}t} & = & 
	-\frac{1}{2}\left[5\mb{Y}^{u}\mb{Y}^{u^{\dagger}}+3\Tr 
	\mb{Y}^{u}\mb{Y}^{u^{\dagger}} + \mb{Y}^{d}\mb{Y}^{d^{\dagger}} - 
2C^{{\rm 
	u}}_{i}g^{2}_{i}\right] \mb{A}^{u}
	\nonumber \\
	 & - & \left[2\mb{A}^{u}\mb{Y}^{u^{\dagger}}+3\Tr 
	\mb{A}^{u}\mb{Y}^{u^{\dagger}} + \mb{A}^{d}\mb{Y}^{d^{\dagger}} - 
2C^{{\rm 
	u}}_{i}g^{2}_{i}M_{i}\right] \mb{Y}^{u},
	\nonumber \\
	\frac{{\rm d} \mb{m}^{2}_{\widetilde{Q}_{L}}}{\rm{d}t}  & = & 
	-\frac{1}{2} \left[\left\{\mb{Y}^{u}\mb{Y}^{u^{\dagger}} + 
	\mb{Y}^{d}\mb{Y}^{d^{\dagger}}, \mb{m}^{2}_{\widetilde{Q}_{L}}\right\}+2 
\left( 
	m^{2}_{h_{u}} \mb{Y}^{u}\mb{Y}^{u^{\dagger}} + m^{2}_{h_{d}} 
	\mb{Y}^{d}\mb{Y}^{d^{\dagger}}\right.\right. 
	\nonumber \\
	 & + & 
\left.\left.\mb{Y}^{u}\mb{m}^{2}_{\tilde{u}_{R}}\mb{Y}^{u^{\dagger}}+
	\mb{Y}^{d}\mb{m}^{2}_{\tilde{d}_{R}}\mb{Y}^{d^{\dagger}} 
	+\mb{A}^{u}\mb{A}^{u^{\dagger}} + 
	\mb{A}^{d}\mb{A}^{d^{\dagger}}\right) \right] + 4 
	C^{Q_{L}}_{i}g^{2}_{i}M^{2}_{i},
	\nonumber \\
	\frac{{\rm d} \mb{m}^{2}_{\tilde{u}_{R}}}{\rm{d}t}  & = & 
	-\left[\left\{\mb{Y}^{u}\mb{Y}^{u^{\dagger}}, 
	\mb{m}^{2}_{\tilde{u}_{R}}\right\}+ 2 \left( 
	m^{2}_{h_{u}} \mb{Y}^{u}\mb{Y}^{u^{\dagger}} + 
	\mb{Y}^{u}\mb{m}^{2}_{\tilde{u}_{R}}\mb{Y}^{u^{\dagger}} 
	+\mb{A}^{u}\mb{A}^{u^{\dagger}} \right)\right] 
	\nonumber \\
	&+& 4 
	C^{u_{R}}_{i}g^{2}_{i}M^{2}_{i},
	\nonumber \\
	\frac{{\rm d} m^{2}_{h_{u}}}{\rm{d}t}  & = & 
	-3 \Tr \left[\mb{Y}^{u}\left(\mb{m}^{2}_{\widetilde{Q}_{L}}+ 
\mb{m}^{2}_{\tilde{u}_{R}} 
	\right) \mb{Y}^{u^{\dagger}} + m^{2}_{h_{u}} 
\mb{Y}^{u}\mb{Y}^{u^{\dagger}} + 
	\mb{A}^{u}\mb{A}^{u^{\dagger}} \right] 
	\nonumber \\
	&+& 4 C^{h_{u}}_{i}g^{2}_{i}M^{2}_{i},
	\label{RGE}
\end{eqnarray}
where $t=1/(16 \pi^2) \log (M_G^2/Q^2)$ is related to the running scale $Q^2$. One has also to consider the additional 
set obtained through the replacement $u 
\leftrightarrow 
d$. In eq.~(\ref{RGE}) we have neglected the leptonic contribution. We 
have 
denoted by $M_{i}$ the gaugino masses and 
 by $C_{i}^{R}$ the quadratic casimir eigenvalues of SU(3), 
SU(2) and U(1) for the representation $R$, with the further definition
\begin{equation}
	C^{{\rm u}}_{i}=C^{Q_{L}}_{i} + C^{u_{R}}_{i} + C^{h_{u}}_{i}, \qquad
	C^{{\rm d}}_{i}=C^{Q_{L}}_{i} + C^{d_{R}}_{i} + C^{h_{d}}_{i}.
	\label{casimiri}
\end{equation}
 
At the electroweak scale, we choose a basis such that the Yukawa couplings 
of up-type quarks are flavour-diagonal,
\begin{equation}
	\mb{Y}^{u}=\mb{Y}^{u}_{D},\quad \mb{Y}^{d}=\mb{K}\mb{Y}^{d}_{D},
	\label{diagY}
\end{equation}
where
\begin{eqnarray}
	\mb{m}^{u}_{D} & = & \diag \left(m_{u},m_{c},m_t\right)=
	\frac{v \sin \beta}{\sqrt{2}}
	\mb{Y}^{u}_D,
	\nonumber \\
	\mb{m}^{d}_{D} & = & \diag \left(m_{d},m_{s},m_b\right)=
	\frac{v \cos \beta}{\sqrt{2}}
	\mb{Y}^{d}_D,
	\label{dmass}
\end{eqnarray}
$v=\sqrt{v_1^2+v_2^2}=246$ GeV and $\tan \beta=v_2/v_1$.

In the super-CKM basis, in which the quark-squark-gaugino couplings are 
equal to the quark-quark-gauge boson ones, the squark mass matrices at the 
electroweak scale are given by
\begin{equation}
	\mb{m}^{2}_{\tilde{u}}=\left(
	\begin{array}{lr}
	\mb{m}^{2}_{\tilde{u}_{LL}} & \mb{m}^{2}_{\tilde{u}_{LR}} \\
	\mb{m}^{2^{\dagger}}_{\tilde{u}_{LR}} & \mb{m}^{2}_{\tilde{u}_{RR}}
	\end{array}
	\right),
	\qquad
	\mb{m}^{2}_{\tilde{d}}=\left(
	\begin{array}{lr}
	\mb{m}^{2}_{\tilde{d}_{LL}} & \mb{m}^{2}_{\tilde{d}_{LR}} \\
	\mb{m}^{2^{\dagger}}_{\tilde{d}_{LR}} & \mb{m}^{2}_{\tilde{d}_{RR}}
	\end{array}
	\right),
	\label{sqmass1}
\end{equation}
where
\begin{eqnarray}
	\mb{m}^{2}_{\tilde{u}_{LL}} &=&
	\mb{m}^{2}_{\widetilde{Q}_{L}} + \left(\mb{m}^{u}_{D}\right)^2 + 
	\frac{M_{Z}^{2}}{6}\left(3-4\sin^{2} \theta\right)\cos 2\beta 
	\nonumber \\
	\mb{m}^{2}_{\tilde{u}_{LR}} &=& 
	-\mu \mb{m}^{u}_{D} \cot \beta -\frac{v \sin \beta}{\sqrt{2}}
	\mb{A}^{u}  
	\nonumber \\
	\mb{m}^{2}_{\tilde{u}_{RR}} &=& 
	\mb{m}^{2}_{\tilde{u}_{R}} +
	\left(\mb{m}^{u}_{D}\right)^2 + \frac{2}{3}M_{Z}^{2}\sin^{2}\theta \cos 
		2 \beta
	\nonumber \\
	\mb{m}^{2}_{\tilde{d}_{LL}} &=&
	\mb{K}^{\dagger}\mb{m}^{2}_{\widetilde{Q}_{L}}\mb{K} + 
\left(\mb{m}^{d}_{D}\right)^2 - 
	\frac{M_{Z}^{2}}{6}\left(3-2\sin^{2} \theta\right)\cos 2\beta 
	\nonumber \\
	\mb{m}^{2}_{\tilde{d}_{LR}} &=& 
	-\mu \mb{m}^{d}_{D} \tan \beta -\frac{v \cos \beta}{\sqrt{2}}
	\mb{K}^{\dagger}\mb{A}^{d}  
	\nonumber \\
	\mb{m}^{2}_{\tilde{d}_{RR}} &=& 
	\mb{m}^{2}_{\tilde{d}_{R}} +
	\left(\mb{m}^{u}_{D}\right)^2 - \frac{1}{3}M_{Z}^{2}\sin^{2}\theta \cos 
		2 \beta.
	\label{sqmass2}
\end{eqnarray}

In the low $\tan \beta$ regime, in which the effects of the Yukawa coupling 
of the bottom quark $Y_{b}$ can be neglected, the effect of $Y_t$ in the 
evolution of the soft breaking masses is to induce a mass splitting 
between squarks; however, the mass matrix for up-type squarks remains 
 diagonal
\cite{Bouquet_et_al, Duncan1}. On the other hand, for large $\tan \beta$ 
$Y_b$ gets 
large and sizeable off-diagonal terms are induced by the evolution 
in $\mb{m}^{2}_{\tilde{u}}$. However, it is 
easy to see from eq.~(\ref{RGE}) that the off-diagonal terms between 
$\tilde{c}$ and $\tilde{t}$ are proportional to $K_{23}$, and therefore at 
most of the order of a few percents,  to be 
compared with the value of $\pi/6$ quoted in refs.~\cite{Li_et_al, 
Couture_et_al}. 

\begin{figure}
\centerline{\epsfig{figure=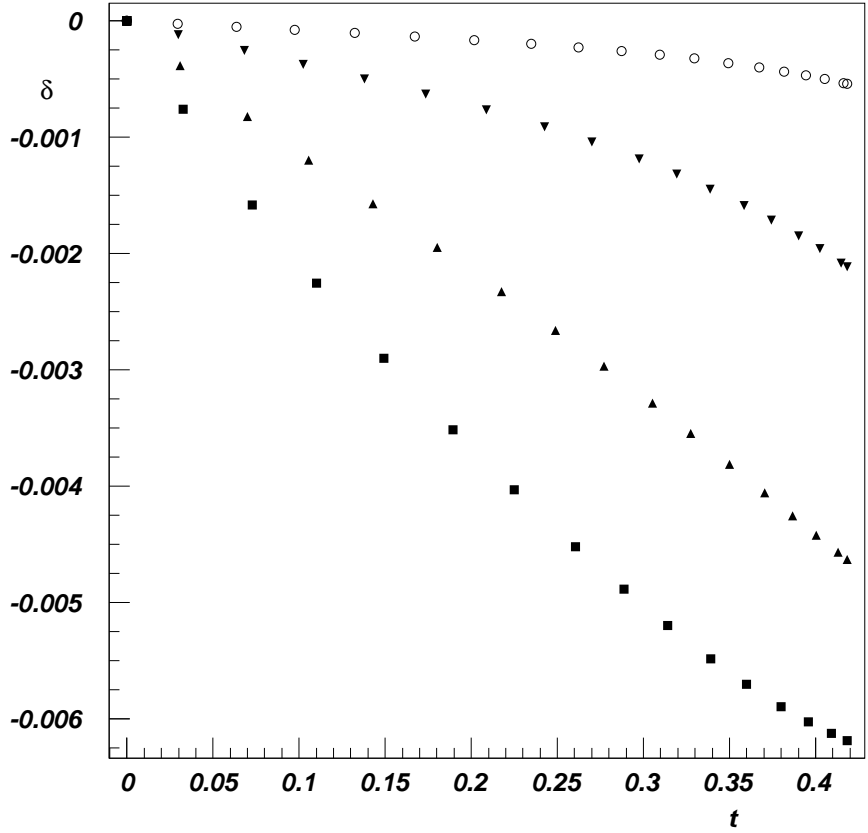,width=1.\linewidth}}
\caption{The renormalization group evolution of the ratio $\delta$ of
the off diagonal element $\tilde{c}_L-\tilde{t}_L$ of the 
squared mass matrix over the average of the diagonal elements of the same 
matrix in the minimal SUSY model with universal coupling 
as a function ot $t=1/(16 \pi^2) \log (M_G^2/Q^2)$. The four curves 
refer to different values of $\tan \beta:~10,~20,~30,~35$ 
(from the upper to the lower).} 
\protect\label{fig:evolution}
\end{figure}

 In figure \ref{fig:evolution} we report for different values of $\tan \beta$
 the renormalization group evolution of the 
value of the ratio of the off-diagonal $\tilde{c}_L - \tilde{t}_L$ 
matrix element squared divided by the average diagonal matrix element squared, which is an estimate of the mixing angle in the mass eigenstate basis. 

As we shall see in section \ref{sec:results}, such a small mixing angle 
renders the prediction of the constrained MSSM for rare top decays 
hard to detect. This is essentially because the only FCNC effects in the squark mass matrix are induced by the quark mass matrix, and are again protected by a GIM mechanism: they disappear when the down quark masses are degenerate.

By relaxing the universality constraints one introduces a
substantial flavour mixing in the theory, resulting in large 
contributions to FCNC processes 
\cite{Gabbiani_and_Masiero}-\cite{Gabbiani_et_al}. There are strong 
constraints on non-universal soft breaking terms involving the first two 
families; however, the off-diagonal squark mass terms between $\tilde{c}$ 
and $\tilde{t}$ are unconstrained by available data on low-energy FCNC 
processes\footnote{In principle, constraints on these mass terms could be 
obtained from a study of chargino contributions to $b \to s \gamma$ decay 
in non-universal SUSY \cite{Pokorski}.}. This means that we 
can envisage a situation in which there is a large mixing angle between    
$\tilde{c}$ and $\tilde{t}$, resulting in large constributions to rare top 
decays, as we shall see in section \ref{sec:results}, in which we 
consider generalizations of the MSSM with arbitrary mixing angles in this 
sector and nonuniversal gaugino masses.

\section{The calculation} 

\begin{figure}
\centerline{\epsfig{figure=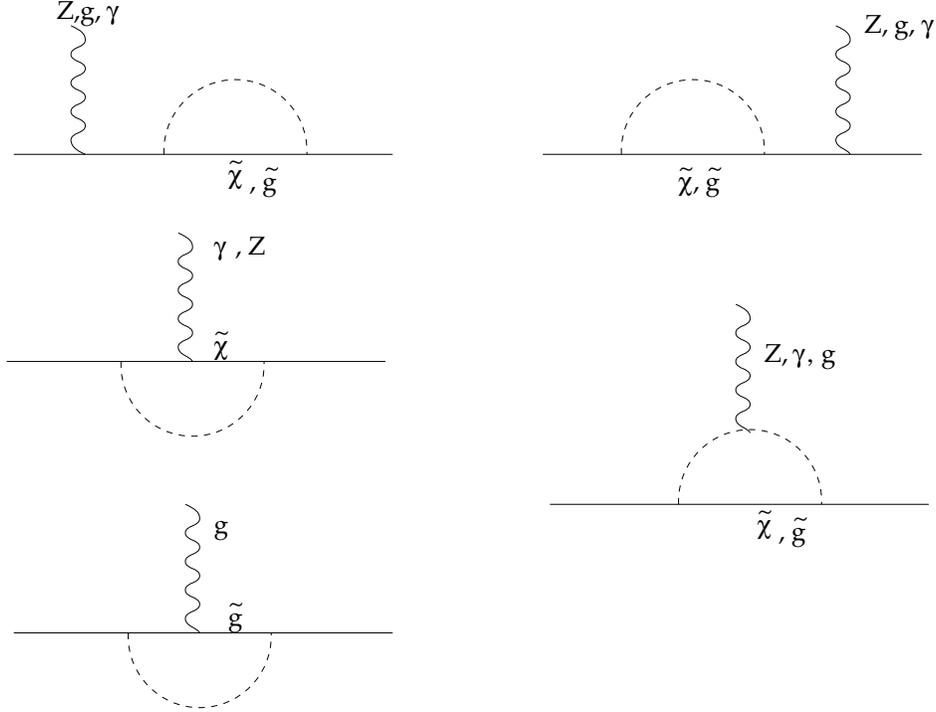,width=.9\linewidth}}
\caption{Diagrams for the processes 
$t \rightarrow c Z^0$, $t \rightarrow c g$, $t \rightarrow c \gamma$. Dashed lines
are for squarks.}
\label{fig:diagrams}
\end{figure}

The calculation was rederived independently.
\\
The general diagrams contributing to top decays are given in figure \ref{fig:diagrams}
for the processes 
$t \rightarrow c Z^0$, $t \rightarrow c g$, $t \rightarrow c \gamma$.

When the flavour changing is a small effect, its contribution can be
safely estimated in the ``mass insertion'' approximation 
\cite{Hall_et_al}-\cite{Gabbiani_et_al},
where the vertices are kept flavour conserving and the flavour off diagonal
terms are treated perturbatively as mass insertions in the
squark propagators. This approximation allows to make estimates which can 
be 
easily transferred from one model to another,
but in general cannot be trusted when the perturbation is not small.\\
This applies in particular to the left-right top squark mixing: 
its main effect is to split considerably the corresponding 
mass eigenstates and to
make the transition less affected by the GIM mechanism.

In general, there are three form factors, 
current conservation restricts them to only two for off shell photons 
and gluons and one for the on shell case. 
For the on shell $Z^0$ case, only two form factors survive.
The contribution from the third form factor, 
due to axial current non conservation when quark masses are different from 
zero,
vanishes when the $Z^0$ is on shell. We have set to zero the charm mass.\\
With respect to previous calculations,
 we take into account the effect of a large left-right stop mixing
which produces a large mass splitting and makes the transitions with a
gluino helicity flip possible and large when the SUSY breaking scale
increases.

In the appendix we report the full result for the form factors relevant 
to the
three different decays for the gluino mediated penguins, which 
represent the dominant contribution when the branching ratios are large. 

\section{The results} 
\label{sec:results}
In table \ref{tab:mssm} we report the results for the constrained MSSM for two values of $\tan \beta$ and one chargino masse near the experimental limit,
i.e around $90~ GeV$. Squark and gluino masses are set between 200 and $300~ GeV$.
\\
Given the smallness of the mixing angle, the correct result for this case 
can be obtained
already in the mass insertion approximation. For moderate values of $\tan \beta$
the chargino contribution dominates: the gluino contribution becomes
comparable only for $\tan \beta = 35$.
In any case the effect is not visible in the universal SUSY 
model.
\begin{table}[h]
\begin{center} 
\begin {tabular} {|| l | c  c | c  c || }
\hline 
& \multicolumn{2}{  c | }{{ $\tan \beta = 2$}}  &
  \multicolumn{2}{ c ||}{{ $\tan \beta = 35$}}   \\
   &  gluino   &   chargino &  gluino   &   chargino    \\
$BR(t\rightarrow c g)$       
& $ 10^{-15}$ & $ 10^{-11}$  & $ 10^{-11}$ & $10^{-12}$          \\
$BR(t\rightarrow c \gamma)$     & $ 10^{-17}$ & $ 10^{-12}$   & $ 10^{-13}$ & $ 10^{-13}$         \\
$BR(t\rightarrow c Z^0)$        & $ 10^{-17}$ & $ 10^{-12}$  & $ 10^{-13}$ & $ 10^{-13}$         \\
\hline
\end{tabular}
\caption{The results of the branching ratios for the MSSM 
with universal soft breaking. }
\vspace{0.3 cm}
\protect\label{tab:mssm}
\end{center}
\end{table}

By abandoning the hypothesis of flavour universality of the soft SUSY
breaking terms, one can explore the result for sizeable values of the
mixing angles.
We have restricted our analysis to the case with a mixing between the
second and the third generation only.
The most general rotation in this four dimensional space (scharm and 
stop, 
both left and right) is  parametrized by six angles that we chose as
the rotation angles in each of the six possible rotation planes.
More explicitely, our rotation matrix $\mb{R}$ is given by:
\begin{equation}
\mb{R}=\mb{R}^{12}\mb{R}^{13}\mb{R}^{14}\mb{R}^{23}\mb{R}^{24}\mb{R}^{34},
\end{equation}
where
\begin{equation}
\mb{R}^{12}=\left(
\begin{array}{cccc}
\cos \theta_{12}& \sin \theta_{12} & 0 & 0\\
-\sin \theta_{12}& \cos \theta_{12} &0 & 0\\
0& 0& 1& 0\\
0& 0 & 0 & 1
\end{array}
\right),
\end{equation}
and similarly for $\mb{R}^{13}$, $\mb{R}^{14}$, $\mb{R}^{23}$, $\mb{R}^{24}$ and $\mb{R}^{34}$.
We generate uniformly in the six dimensional space of rotation angles a mixing
matrix in the basis where the squark mass matrix is diagonal with
definite eigenvalues.
Our default choice are the values at the lower experimental bounds:
\\
\begin{eqnarray}
 &  m_{\tilde{c}_1} = 220 ~GeV, 
 &  m_{\tilde{c}_2} = 260 ~GeV, \nonumber \\
 &  m_{\tilde{c}_1} = 90 ~GeV, 
 &  m_{\tilde{c}_2} = 180 ~GeV, \nonumber 
\end{eqnarray}
where the labels $1$ and $2$ denote the proper eigenstates of the mass matrix.
The gluino mass has been set equal to $154~GeV$.
\\
For each choice of the matrix we calculate the branching ratio.

\begin{figure}
\centerline{\epsfig{figure=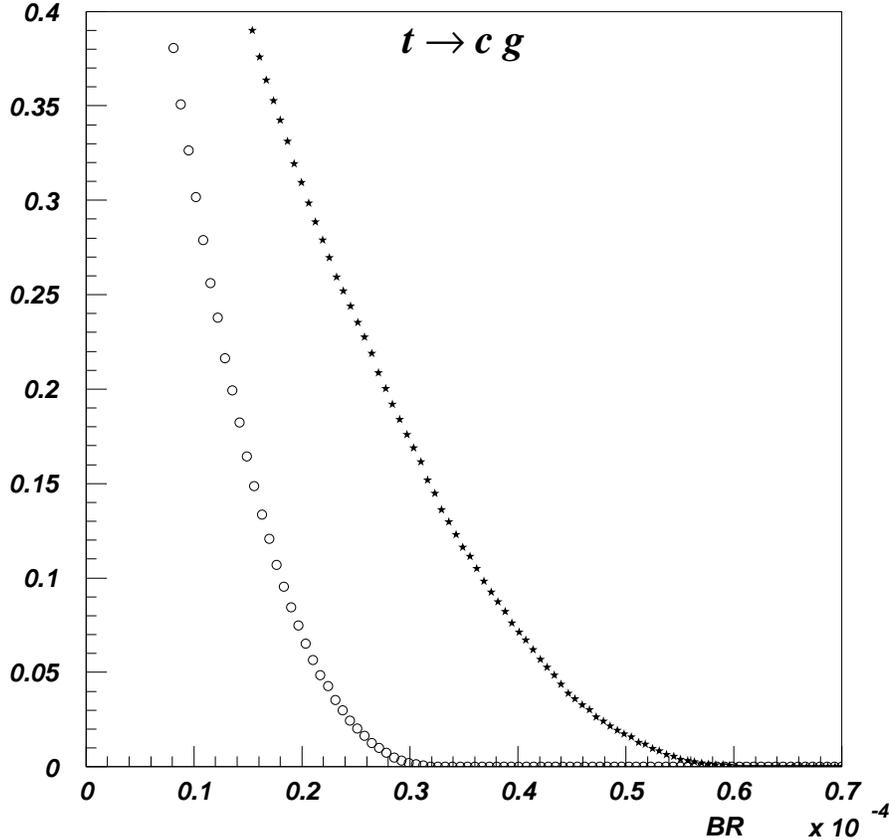,width=1.\linewidth}}
\caption{The integral of the events with branching ratio greater than $BR$
generated with random rotation matrix for the $t \to c g$ decay with 
gluino penguins for the default choice of squark mass eigenvalues (upper curve)
and for a choice with a smaller splitting among the eigenstates (lower curve).}
\protect\label{fig:gluon}
\end{figure}

In figure \ref{fig:gluon} we plot for the gluon decay with
intermediate gluino exchange the percentage of the 
cases where the resulting branching ratio is bigger than a given value.
Such a percentage gives an indication about the naturalness of the
corresponding values of the branching ratio.
The maximum value of the branching ratio is of the order of $5.~10^{-5}$ and a reasonable percentage 
of the order of $20\%$ is
obtained for branching ratios larger than
 $3.~10^{-5}$, i.e. for a detectable signal.
On the same figure is given the case where the eigenvalues of stop states have been changed from
90 and 180 to 120 and 160.
 The curve becomes steeper because of the more severe 
cancellations due to the GIM mechanism.

\begin{figure}
\centerline{\epsfig{figure=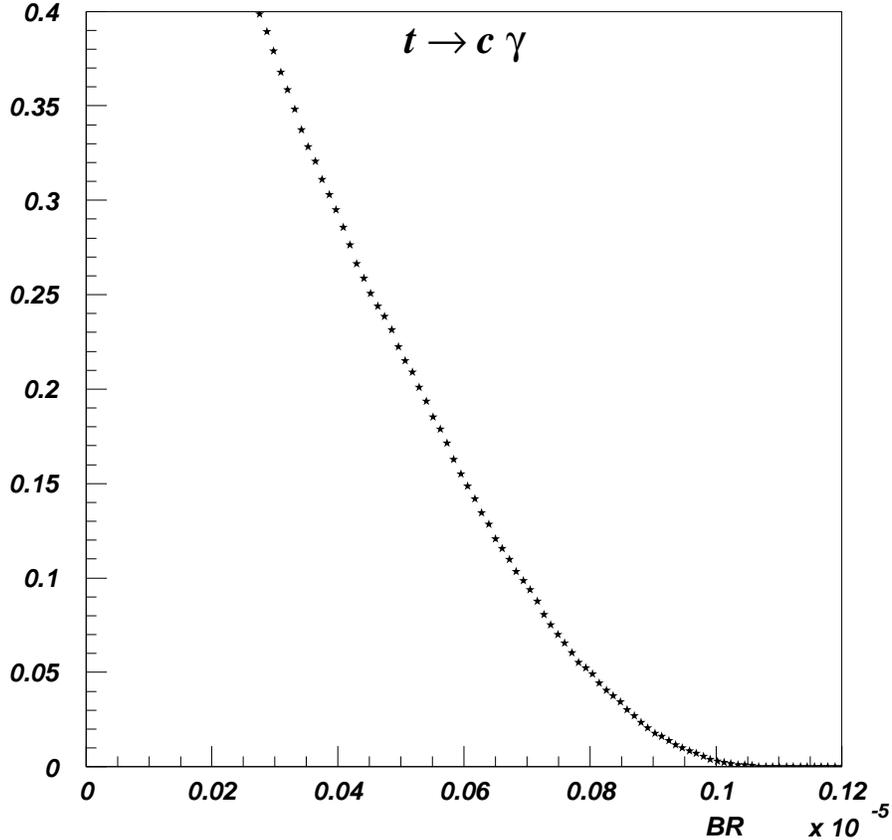,width=1.\linewidth}}
\caption{The same as in figure 2 for the $t \to c \gamma$ decay 
and the default choice of the mass eigenvalues.}
\protect\label{fig:photon}
\end{figure}

\begin{figure}
\centerline{\epsfig{figure=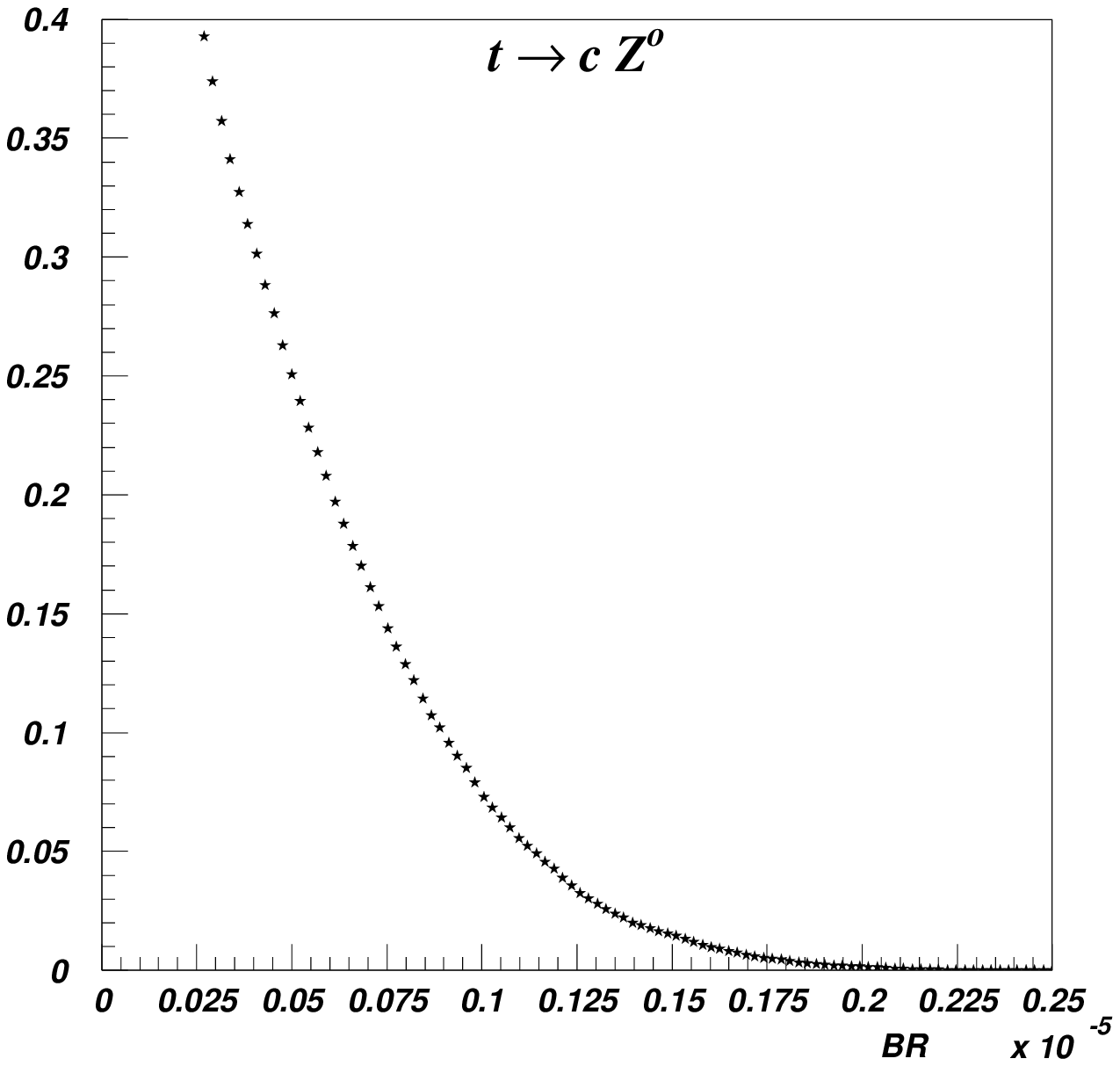,width=1.\linewidth}}
\caption{The same as in figure 2 for the $t \to c Z^0$ decay 
and the default choice of the mass eigenvalues.}
\protect\label{fig:z0}
\end{figure}

Figures \ref{fig:photon} and \ref{fig:z0} refer to the photon and $Z^0$ case, respectively.
The values are smaller by a factor ranging from  20 to 50
coming basically from the different colour charge.

\begin{figure}
\centerline{\epsfig{figure=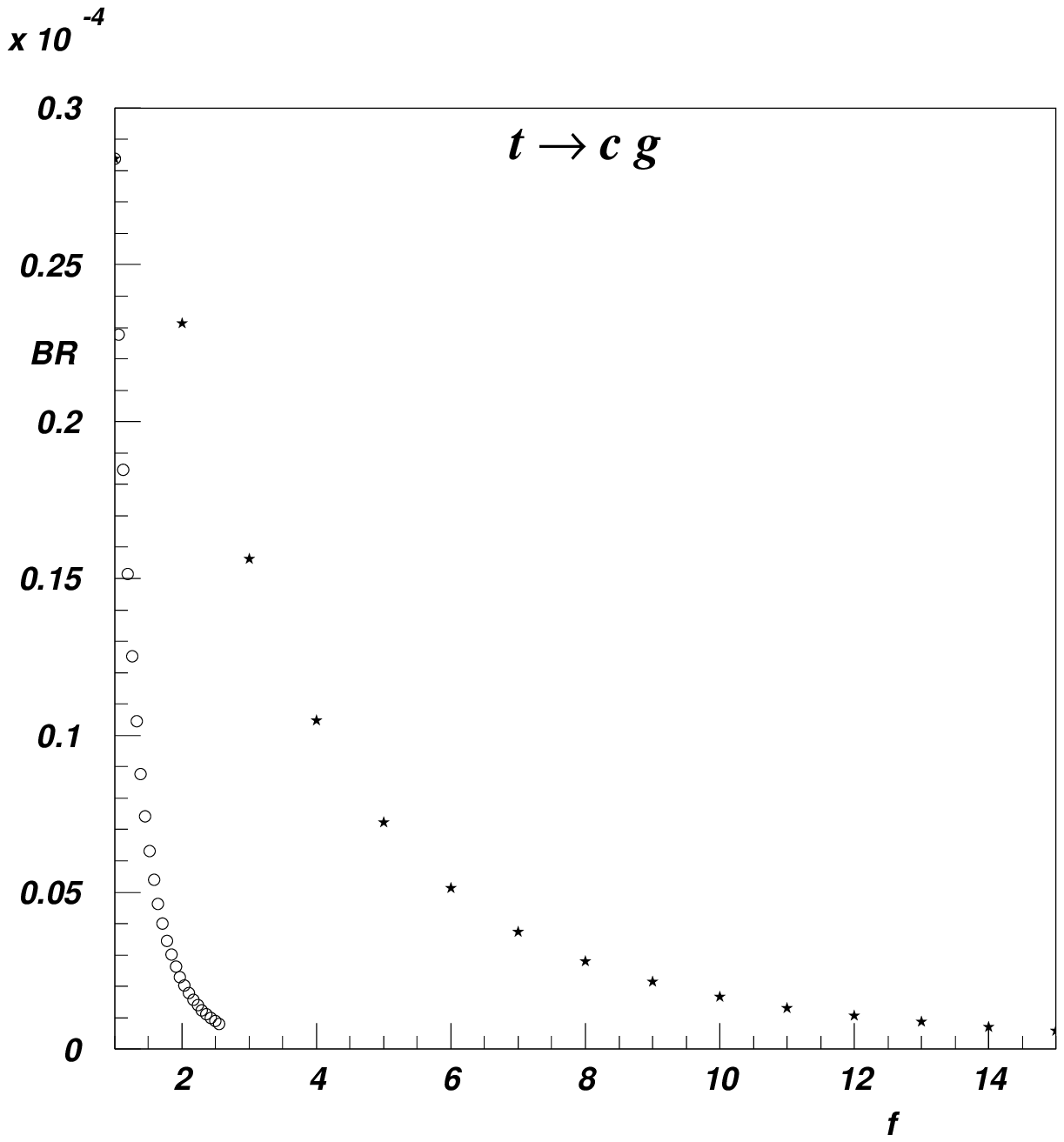,width=1.\linewidth}}
\caption{The variation of the branching ratio for a particular and 
favourable choice of the mixing matrix with  the 
squark (upper curve) or the gluino (lower curve) mass rescaling 
by a factor $f$. }
\protect\label{fig:factor}
\end{figure}

In figure \ref{fig:factor} we report the variation,
 for a fixed favourable choice of
the mixing angles, of the branching ratio 
with the values of the squark or of the gluino masses, rescaled
 by a factor $f$ with respect to their reference values.
 The dependence upon the gluino mass at fixed default values 
of the squark masses appears to be more
relevant than the reverse case.

For each rotation matrix and choice of the mass eigenvalues we can
reconstruct the original non diagonal mass matrix 
and in particular we can analyse which
entries of this matrix should be large to reach the highest values of
the branching ratio.
This can be seen by plotting, for the events
generated with a random rotation matrix
where the branching ratio is higher than a given threshold value,
 the correlation
of two off diagonal entries of the mass matrix 
normalized to the average of the diagonal entries.

\begin{figure}[p]
\centerline{\epsfig{figure=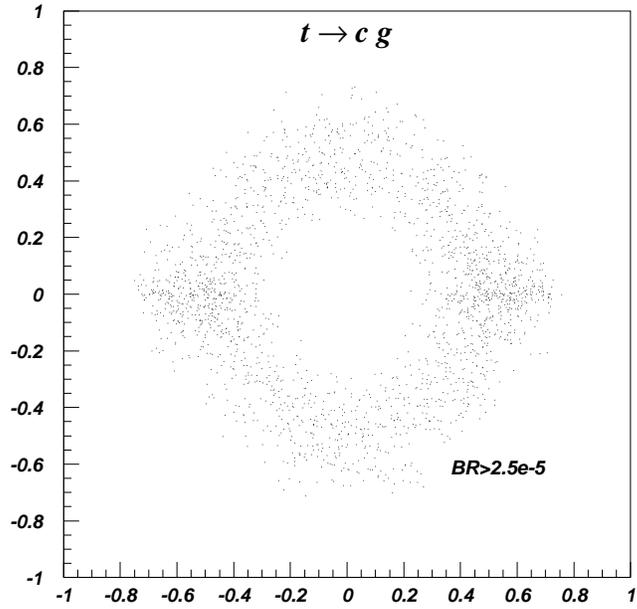,width=.75\linewidth}}
\centerline{\epsfig{figure=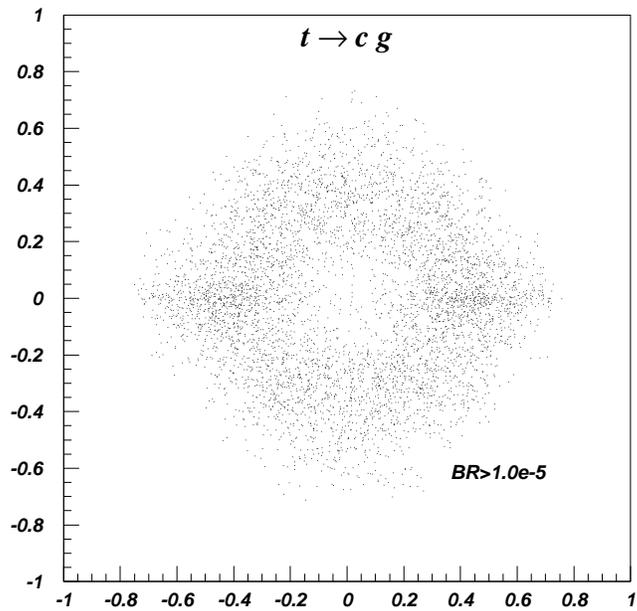,width=.75\linewidth}}
\caption{The correlation for the events generated with random rotation 
matrices of the matrix elements of the original off diagonal squared mass 
matrix $\tilde{c}_L-\tilde{t}_R$ and $\tilde{c}_R-\tilde{t}_L$ for 
two different cuts of the BR.}
\protect\label{fig:correlations}
\end{figure}
In figures \ref{fig:correlations} we plot the correlation of $\tilde{c}_L-\tilde{t}_R$ 
versus $\tilde{c}_R-\tilde{t}_L$ for 
two different cuts on the branching ratio. Among all
possible correlations, only this correlation shows 
a hole in the center (when both values are close to zero)
which increases with the value
of the cut. The condition that at least one 
of the two off diagonal entries
$\tilde{c}_L-\tilde{t}_R$ or $\tilde{c}_R-\tilde{t}_L$
should be large is a necessary one when a large branching ratio
is required. This cannot be realized in the minimal SUSY model
with universal couplings.\\

Flavor changing top decays are expected to be visible only in extensions
 of the minimal supersymmetric standard model 
where the soft breaking is flavour dependent and non universal.\\

NOTE: While completing this work we have seen
a paper by Lopez {\it et al.}, hep-ph 9702350, where the same processes were considered.
Their conclusions on the yield of FCNC in SUSY models with universal
breaking are different from ours.

\section*{Appendix}

\def\msa2{m_{A}^2}
\def\msb2{m_{B}^2}
\def\mgg{m_{\tilde g}^2}
\def\mtt{m_{top}^2}
\def\mg{m_{\tilde g}}
\def\mt{m_{top}}
\def\dm{(m_{top}^2-\q2)}
\def\pt2{\mtt}
\def\q2{q^2}

\newcommand{\bb} [3] {  B_0(#1, #2, #3)  }
\newcommand{\cc} [3] {  C_0(#1, #2, #3)  }
\newcommand{\bx} [3] {  B_x(#1, #2, #3)  }
We report in this appendix the result 
for the gluino mediated diagrams.

The effective vertex can be parametrized as follow:
\begin{eqnarray}
 & -i ~\bar{u} (p)  \lbrace&
      P_R (F_L^a q^2 \gamma^{\mu}+ F_L^b \not\! q q^{\mu} + G_L i \sigma^{\mu \nu} q_{\nu}) + \nonumber \\
 & &  P_L (F_R^a q^2 \gamma^{\mu}+ F_R^b \not\! q q^{\mu} + G_R i \sigma^{\mu \nu} q_{\nu}) \rbrace~\epsilon_\mu~ u(p+q)  
\end{eqnarray}
where $P_{L,R} = \frac{(1\mp \gamma_5)}{2}$ and 
$\epsilon_\mu = \epsilon_\mu^a \mb{T}^a$ for gluon and 
$\epsilon_\mu = \epsilon_\mu \uno$ for photon or $Z^0$. In the photon and the gluon case  gauge invariance
implies  $F^a = -F^b$, a relation that we have explicitely
checked.\\ 
The form factors $F^a$ can be written as
\begin{eqnarray}
& & F_L^a =  \frac{ g_s^2} {4 \pi^2} \sum_{A,B=1}^4 \lbrace
                   \mb{R}^\dagger_{\tilde{c}_L,B} \mb{R}_{A,\tilde{t}_L}~F_{1L}^a (A,B)  -
                   \mb{R}^\dagger_{\tilde{c}_L,B} \mb{R}_{A,\tilde{t}_R}~F_{2L}^a (A,B)
              \rbrace \nonumber \\
& & F_R^a = \frac{ g_s^2} {4 \pi^2}\sum_{A,B=1}^4 \lbrace
                     \mb{R}^\dagger_{\tilde{c}_R,B} \mb{R}_{A,\tilde{t}_R}~F_{1R}^a (A,B)  -
                     \mb{R}^\dagger_{\tilde{c}_R,B} \mb{R}_{A,\tilde{t}_L}~F_{2R}^a (A,B)
              \rbrace .\nonumber \\
\end{eqnarray}
Analogous expressions can be written for $F^b$ and $G$.
The matrix $\mb{R}$ diagonalizes the squared squark mass matrix, and 
the indices $A$ and $B$ identify the eigenvalues and the corresponding eigenvectors:
\begin {equation}
\left(\matrix{\tilde{q}_1\cr
              \tilde{q}_2\cr
              \tilde{q}_3\cr 
              \tilde{q}_4\cr }\right)= \mb{R}
\left(\matrix{\tilde{c}_L\cr
              \tilde{c}_R\cr
              \tilde{t}_L\cr 
              \tilde{t}_R\cr}\right).
\end{equation}
In general $f$ can be split in a piece proportional 
 to $C_F=\frac{N^2-1}{2N} =4/3$ and one  
proportional to $C_G=N=3$, the latter is present in the gluon case.
For the $C_G$ terms 
we report  only the expression of the form factor relevant 
for on shell gluon.  
For the $C_F$ terms we report the result at $q^2\not=0$ which is needed 
for the $Z^0$ case. The form factor $F^b$ 
does not contribute to the branching ratio for on shell vector bosons.

The  formulas 
are parametrized in terms of
 the following couplings of the gauge boson $V=g,\gamma,Z^0$ to the left or right up quarks $a^{qqV}_{L(R)}$ and of
 the couplings of the gauge boson to the up squarks in the mass eigenstates basis
$a_{\tilde{q}\tilde{q}V}(A,B)$:\\
\begin{eqnarray}
& \mbox{for the gluon } & 
a^{qqg}_L= a^{qqg}_R= -g_s, \nonumber \\ 
& \mbox{for the photon } & a^{qq\gamma}_L= a^{qq\gamma}_R= -g_{em} \frac{2}{3},  \nonumber \\ 
& \mbox{and for $Z^0$ } & \nonumber \\ 
& & a^{qqZ}_L= \frac{g_w}{2 \cos \theta _w} (-1 +  \frac{4}{3} \sin^2 \theta _w) \nonumber \\ 
& & a^{qqZ}_R= \frac{g_w}{2 \cos \theta _w} (  \frac{4}{3} \sin^2 \theta _w) .\nonumber  
\end{eqnarray}

The $Z^0$  coupling  is different for left 
and right handed quarks and squarks and therefore is not diagonal 
in the squark mass eigenstates basis, 
$a_{\tilde{q}\tilde{q}Z}(A,B) = \left (\mb{R} ~\hat{a}_{\tilde{q}\tilde{q}Z}~\mb{R}^\dagger \right )_{BA}$ where 
\begin{displaymath}
\hat{a}_{\tilde{q}\tilde{q}Z} = \mbox{diag} \left (a^{qqZ}_L, 
a^{qqZ}_R, 
a^{qqZ}_L,
a^{qqZ}_R\right). 
\end{displaymath}
We have used the Feynman rules described in \cite{Haber_and_Kane}.
The results have been obtained with algebraic package FORM\cite{form}.

\noindent For the photon or $Z^0$ $F^a_{..}(.,.)= F^a_{..}(.,.)^{C_F}$ and 
$G_{..}(.,.)= G_{..}(.,.)^{C_F}$
%
%
{\footnotesize
\begin{eqnarray}
 & &  F_{1L(R)}^a(A,B)^{C_F} = \nonumber \\
 & &  a^{qqV}_{L(R)}\; \delta_{A,B} ~C_F \cdot \nonumber \\
 & &  {{1} \over {2 \q2 }}  \lbrace {  \bx{\pt2}{\mgg}{\msa2}+\bb{\pt2}{\msa2}{\mgg} }    \rbrace 
\nonumber \\
 & &  a_{\tilde{q}\tilde{q}V}(A,B)  ~C_F \cdot   \lbrace  \nonumber \\
 & &  +{{\mtt (\msb2-\mgg)}  \over {\dm^2 \q2}} \lbrace { \bb{\pt2}{\mgg}{\msa2}-\bb{0}{\mgg}{\msb2}
                                          -(\msa2-\msb2) \cc{\msa2}{\mgg}{\msb2}}    \rbrace  
\nonumber \\
 & &  +{{3}\over{2}}{{\mtt (\msb2-\mgg)}  \over {\dm^3 }} \lbrace { \bb{\pt2}{\mgg}{\msa2}-\bb{\q2}{\msb2}{\msa2}
                                           +(\msb2-\mgg) \cc{\msa2}{\mgg}{\msb2}}    \rbrace 
\nonumber \\
 & &  -{{1}\over{2}}{{\mtt} \over {\dm \q2 }} \lbrace { \bb{\pt2}{\mgg}{\msa2}+\bx{\pt2}{\msa2}{\mgg}}
                                              \rbrace 
\nonumber \\
 & &  -{{1}\over{2}}{{(\msb2 -\mgg)} \over {\dm^2 }}  \lbrace { \bb{\pt2}{\mgg}{\msa2}-\bb{\q2}{\msa2}{\msb2}
                                           +(\msb2 -\mgg) \cc{\msa2}{\mgg}{\msb2}}    \rbrace 
\nonumber \\
 & &  -{{1}\over{2}}{{\mtt} \over {\dm^2 }}  \lbrace { \bx{\pt2}{\msa2}{\mgg}-\bx{\q2}{\msa2}{\msb2}-2(\msb2 -\mgg) \cc{\msa2}{\mgg}{\msb2}}
                                              \rbrace 
\nonumber \\
 & &  -{{1}\over{2}}{{1} \over {\dm }}  \lbrace { 
\bx{\q2}{\msa2}{\msb2}
                                           -\mgg
 \cc{\msa2}{\mgg}{\msb2}} -\frac{1}{2}   \rbrace 
\nonumber \\
 & &  -{{1}\over{2}} {{ (\msb2-\mgg)}  \over {\dm \q2}} \lbrace { \bb{\pt2}{\mgg}{\msa2}-\bb{0}{\mgg}{\msb2}  -(\msa2-\msb2) \cc{\msa2}{\mgg}{\msb2}}    \rbrace  \rbrace  
\nonumber \\
%
%
 & &    \\
 & &  F_{2L(R)}^a(A,B)^{C_F} = \nonumber \\
 & &  a^{qqV}_{L(R)}\; \delta_{A,B} ~C_F \cdot \nonumber \\
 & &   {{1}\over{2}}{{\mt \mg} \over {\mtt \q2 }} \lbrace {  \bb{\pt2}{\mgg}{\msa2}- \bb{0}{\msa2}{\mgg}}
                                              \rbrace \rbrace
\nonumber \\
 & &  a_{\tilde{q}\tilde{q}V}(A,B) ~C_F \cdot   \lbrace  \nonumber \\
 & &  -{{1}\over{2}}{{\mt \mg}  \over {\dm \q2}} \lbrace { \bb{\pt2}{\mgg}{\msa2}-\bb{0}{\mgg}{\msb2}
                                          -(\msa2-\msb2) \cc{\msa2}{\mgg}{\msb2}}    \rbrace  
\nonumber \\
 & &  -{{\mt \mg }  \over {\dm^2 }} \lbrace { \bb{\pt2}{\mgg}{\msa2}-\bb{\q2}{\msb2}{\msa2}
                                           +(\msb2-\mgg) \cc{\msa2}{\mgg}{\msb2}}    \rbrace 
\nonumber \\
 & &  -{{1}\over{2}}{{\mt \mg} \over {\dm }} \lbrace {  
\cc{\msa2}{\mgg}{\msb2}}    \rbrace\rbrace 
\end{eqnarray}
}
It can be verified that the above expression 
has a smooth limit when $\q2 \to 0$.
%
%
{\footnotesize
\begin{eqnarray}
 & &  G_{1L}(A,B)^{C_F}  = G_{1R}(A,B)^{C_F}  = \nonumber \\
 & &  a_{\tilde{q}\tilde{q}V}(A,B) ~C_F \cdot   \lbrace  \nonumber \\
 & &  +{\mt}{{(\msb2-\mgg)}  \over {\dm^2 }} \lbrace { \bb{\pt2}{\mgg}{\msa2}-\bb{0}{\mgg}{\msb2}-(\msa2-\msb2) \cc{\msa2}{\mgg}{\msb2}}\rbrace  
\nonumber \\
 & & +{{3}\over{2}}{{\q2 \mt (\msb2-\mgg)}  \over {\dm^3 }} \lbrace { \bb{\pt2}{\mgg}{\msa2}-\bb{\q2}{\msb2}{\msa2}                                           +(\msb2-\mgg) \cc{\msa2}{\mgg}{\msb2}}    \rbrace 
\nonumber \\
 & &  -{{1}\over{2}}{{\mt \q2} \over {\dm^2}} \lbrace { \bx{\pt2}{\msa2}{\mgg}-\bx{\q2}{\msa2}{\msb2}
                                           -2(\msb2-\mgg) \cc{\msa2}{\mgg}{\msb2}}    \rbrace 
\nonumber \\
 & &  -{{1}\over{2}}{{\mt} \over {\dm}} \lbrace { \bb{\pt2}{\msa2}{\mgg}+2\bx{\pt2}{\msa2}{\mgg}
 -\msb2 \cc{\msa2}{\mgg}{\msb2}} -\frac{1}{2}   \rbrace  \rbrace 
\nonumber \\
 \\
%
%
 & &  G_{2L}(A,B)^{C_F}  = G_{2R}(A,B)^{C_F}  = \nonumber \\
 & &  a_{\tilde{q}\tilde{q}V}(A,B)  ~C_F \cdot   \lbrace  \nonumber \\
 & &  {-}{{1}\over{2}}{{ \mg}  \over {\dm }} \lbrace { \bb{\pt2}{\mgg}{\msa2}-\bb{0}{\mgg}{\msb2}
                                          -(\msa2-\msb2) \cc{\msa2}{\mgg}{\msb2}}    \rbrace  
\nonumber \\
 & &  - {{\mg \q2 }  \over {\dm^2 }} \lbrace { \bb{\pt2}{\mgg}{\msa2}-\bb{\q2}{\msb2}{\msa2}
                                           +(\msb2-\mgg) \cc{\msa2}{\mgg}{\msb2}}    \rbrace 
\nonumber \\
 & & -{{1}\over{2}}{{\q2 \mg} \over {\dm}} \lbrace {   \cc{\msa2}{\mgg}{\msb2}}                                              \rbrace \rbrace 
\end{eqnarray}}

\noindent For the gluon $G_{..}(.,.)= G_{..}(.,.)^{C_F}+G_{..}(.,.)^{C_G}$
{\footnotesize
%
%
\begin{eqnarray}
 & &   G_{1L}(A,B)^{C_G}  = G_{1R}(A,B)^{C_G}  = \nonumber \\
 & &  -g_s ~\delta_{A,B} ~\frac{1}{2}C_G \cdot   \lbrace  \nonumber \\
 & &  -{\mt}{{(\msb2-\mgg)}  \over {(\mtt)^2 }} \lbrace { \bb{\pt2}{\mgg}{\msa2}-\bb{0}{\mgg}{\msb2}-(\msa2-\msb2) \cc{\msa2}{\mgg}{\msb2}}\rbrace   
\nonumber \\
 & & +{{1}\over{2}}{{ \mt }  \over {\mtt }} \lbrace { \bb{\pt2}{\mgg}{\msa2}+2\bx{\pt2}{\msa2}{\mgg}                                           -\msb2 \cc{\msa2}{\mgg}{\msb2}}  -\frac{1}{2}  \rbrace  
\nonumber \\
 & & +{\mt}{{(\msa2-\mgg)}  \over {(\mtt)^2 }}\lbrace { \bb{\pt2}{\mgg}{\msa2}-\bb{0}{\mgg}{\msa2}}\rbrace  
\nonumber \\
 & & +{{1}\over{2}}{{ \mt }  \over {\mtt }} \lbrace { \bb{\pt2}{\mgg}{\msa2}+2\bx{\pt2}{\mgg}{\msa2}                                           -\mgg \cc{\mgg}{\msa2}{\mgg}}   -\frac{1}{2} \rbrace \rbrace 
 \\
\nonumber \\
%
%
 & & G_{2L}(A,B)^{C_G}  = G_{2R}(A,B)^{C_G}  = \nonumber \\
 & &  -g_s ~\delta_{A,B}   ~\frac{1}{2}C_G\cdot   \lbrace  \nonumber \\
 & & +{{1}\over{2}}{{ \mg}  \over {\mtt }} \lbrace { \bb{\pt2}{\mgg}{\msa2}-\bb{0}{\mgg}{\msb2}
                                          -(\msa2-\msb2) \cc{\msa2}{\mgg}{\msb2}}    \rbrace    
\nonumber \\
 & & -{{1}\over{2}}{{ \mg}  \over {\mtt }} \lbrace { \bb{\pt2}{\mgg}{\msa2}-\bb{0}{\mgg}{\msa2}
                                          +\mtt \cc{\mgg}{\msa2}{\mgg}}    \rbrace \rbrace  
\end{eqnarray}
}

Where:
\begin{eqnarray}
 & &  \!\!\!\!\!\! \bb{p^2}{m_1^2}{m_2^2} = \mbox{pole terms +} \nonumber \\
 & &  - \int_0^1 d\alpha \log~
                    ( p^2 \alpha^2 +(m_1^2 -m_2^2 - p^2) \alpha + m_2^2) 
\nonumber \\
 & &  \!\!\!\!\!\! \bx{p^2}{m_1^2}{m_2^2} =  \mbox{pole terms +} \nonumber \\
 & &  + \int_0^1 d\alpha~\alpha~\log~
                    ( p^2 \alpha^2 +(m_1^2 -m_2^2 - p^2) \alpha + m_2^2) 
\nonumber \\
 & &  \!\!\!\!\!\! \cc{m_1^2}{m_2^2}{m_3^2} = \nonumber \\
 & & \int_0^1 d\alpha
      \frac{1}{\dm \alpha + (m_3^2 -m_2^2) }
                     \log\frac
                    {( \mtt \alpha^2 +(m_1^2 -m_2^2 - \mtt) \alpha + m_2^2)}
                    {( \q2  \alpha^2 +(m_1^2 -m_3^2 - \q2 ) \alpha + m_3^2)}
\nonumber \\
\end{eqnarray}

The above integrals contain divergent parts (pole terms) which cancel in the finite result because the matrix \mb{R} is unitary.\\
The integrals are performed in the dimensional regularization scheme:
in particular the reduction of tensor integrals to scalar integrals 
is performed in $4-\epsilon$ dimensions, while the gamma matrices are 
kept in 4 dimension. The final result being finite is indipendent 
upon the particular regularization that we have chosen. 

In terms of the above form factors the expression 
for the branching ratios reads:  
\begin{displaymath}
 BR = \frac{ \Gamma (t \to c + V)}{\Gamma (t \to b + W)}
\mbox{\hspace{0.8 cm} with }\Gamma (t \to b + W)\sim 1.52~GeV
\mbox{ and }\end{displaymath}
\begin{eqnarray}
&\Gamma (t \to c + \gamma) =  & \frac{1}{16 \pi}   m_{top}^3 
\left( G_L^2+G_R^2\right) |_{(q^2=0)}
\nonumber \\
&\Gamma (t \to c + g) =  & \frac{1}{16 \pi} C_F  m_{top}^3 
\left( G_L^2+G_R^2\right) |_{(q^2=0)}
\nonumber \\
& \Gamma (t \to c + Z^0) = & \frac{1}{32 \pi} \frac{m_{top}^2-m_Z^2}{2 m_{top}^3} \cdot\nonumber \\
& & 
\left\{ \left( {F^a_L}^2+{F^a_R}^2  \right) |_{(q^2=m_Z^2)}
        \left( 2 m_{top}^4 m_Z^2+ 2 m_{top}^2 m_Z^4 - 4  m_Z^6 \right)\right.\nonumber \\
& & + \left( {F^a_L}G_L+{F^a_R}G_R  \right) |_{(q^2=m_Z^2)} 
        \left(-12 m_{top}^3 m_Z^2+ 12 m_{top} m_Z^4 \right)\nonumber \\
& & + \left(  G_L^2+G_R^2  \right) |_{(q^2=m_Z^2)}
        \left( 4 m_{top}^4 - 2 m_{top}^2 m_Z^2 - 2  m_Z^4 \left.\right)
\right\} \nonumber \\\end{eqnarray}

\def\NPB #1 #2 #3 {Nucl.~Phys.~{\bf#1} (#2)\ #3}
\def\PRD #1 #2 #3  {Phys.~Rev.~{\bf#1} (#2)\ #3}
\def\PLB #1 #2 #3 {Phys.~Lett.~{\bf#1} (#2)\ #3}
\def\PRL #1 #2 #3 {Phys.~Rev.~Lett.~{\bf#1} (#2)\ #3}
\def\PR  #1 #2 #3 {Phys.~Rep.~{\bf#1} (#2)\ #3}

\def\etal{{\it et al.}}
\def\ibid{{\it ibid}.}

\end{document}